# The structure of the superconducting gap in $MgB_2$ from point-contact spectroscopy

Y.Bugoslavsky*, Y.Miyoshi, G.K.Perkins, A.V.Berenov, Z.Lockman, J.L. MacManus-Driscoll, L.F.Cohen, A.D.Caplin,
*Centre for High-Temperature Superconductivity, Imperial College, London, SW7 2BZ, UK*
* *General Physics Institute, Moscow 117942, Russia*
H Y Zhai, M P Paranthaman and H M Christen
*Oak Ridge National Laboratory, Oak Ridge, TN 37931-6056, USA*
M. Blamire
*Deptartment of Materials Science, Cambridge University, Pembroke St., Cambridge, CB2 3QZ*

## Abstract

We have studied the structure of the superconducting gap in $MgB_2$ thin films by means of point-contact spectroscopy using a gold tip. The films were produced by depositing pure boron on a sapphire substrate, using e-beam evaporation, followed by reaction with magnesium vapour. The films have a $T_c$ of 38.6 ± 0.3 K and resistivity of about 20 μOhm·cm at 40 K. The point-contact spectra prove directly the existence of a multi-valued order parameter in $MgB_2$, with two distinct values of the gap, $\Delta_1$=2.3±0.3 meV and $\Delta_2$=6.2±0.7 meV at 4.2 K. Analysis of the spectra in terms of the Blonder-Tinkham-Klapwijk model reveals that both gaps close simultaneously at the $T_c$ of the film. Possible mechanisms that can explain the intrinsic co-existence of two values of the gap are discussed.

## Introduction

Since the recent discovery of superconductivity in magnesium diboride [1], there has been impressively rapid progress in understanding both its basic properties and possible applications [2]. However there still remain open questions concerning the structure of superconducting state and the underlying pairing mechanism. Reliable experimental evidence on the order parameter (or, equivalently, the superconducting gap, Δ) is the key issue for understanding superconductivity in $MgB_2$. Throughout the several months of intensive studies of $MgB_2$, the value of the gap has been the subject of substantial controversy. The earliest experiments yielded observations of distinct gap features in tunneling spectra [3, 4, 5, 6] (obtained with scanning tunneling microscopy (STM) or by point-contact techniques), but the values reported by different groups were at odds with each other, ranging from 1.5 meV to 7.5 meV. The classical Bardeen-Cooper-Schrieffer (BCS) theory directly relates Δ to $T_c$; the ratio is $2\Delta/k_BT_c$=3.52. Had $MgB_2$ been a simple BCS superconductor with $T_c$=39 K, its gap should have been Δ=5.9 meV. Higher values inferred experimentally could be explainable in terms of strong coupling, but no regime in the BCS framework would be compatible with a small Δ in a high-$T_c$ material. Therefore it was natural to assume that the low-gap data arose from the surface layer of the samples, which could have a degraded $T_c$.

Several groups have observed double gap structures, with the smaller gap in the range 2-3 meV, and the larger one at 5-8 meV. The results of Giubileo *et al.* (STM) [7], and Szabo *et al.* (point contact) [8] suggest that both gaps are intrinsic, and disappear simultaneously at $T_c$. On the other hand, Plecenik et al., ($MgB_2$ wire, Ag or In drops used to form N-S contacts) [9] and Carapella *et al.* (planar junction) [10] interpret their results in terms of two superconducting phases with different $T_c$'s (with the lower $T_c$ around 15-20 K). The spectroscopic techniques used in these experiments are in principle similar, yet there is little consistency between the results, particularly concerning the relative amplitudes of the features that are taken as the evidence for two distinct gaps. For example, two clear sets of pronounced peaks were observed in ref. [8], whereas the large gap feature is only 10-20% of the amplitude of the primary low-gap feature in ref. [7]. Additionally in refs. [9] and [10], the authors infer a double-gap structure from fitting of the data only, whereas the data themselves lack any clear evidence of two features. To add to the confusion, it is implied in both [7] and [8] that the double-gap spectra were not always obtained; sometimes only the lower gap was observed. Similarly, Gonnelli *et al.*[11], fitted their point-contact data (platinum and gold tips) to just a single-gap function and obtained excellent agreement; their values of Δ are in the range between 2.4 and 3.2 meV. It appears therefore that the results are very much dependent on the sample surface quality and homogeneity, which raises the issue of to what extent these measurements represent the intrinsic properties of $MgB_2$. Another dimension to the problem is brought about by the fact that $MgB_2$ is an anisotropic material [12]; gap measurements can therefore be sensitive to the sample texture as well. Indeed, Senor et al., [13] found different values of $\Delta$ depending upon sample preparation and orientation.

A number of other experiments suggest the presence of a double gap excitation spectrum in $MgB_2$, although in a less direct way. These include specific heat measurements [14], photoemission spectroscopy [15], and Raman spectroscopy [16]. The overall picture that emerges is that there are strong indications of non-trivial superconducting order parameter in $MgB_2$. At the same time a significant number of *direct* measurements yield just a single gap feature (positioned at too low an energy), or a barely discernible contribution of a higher-energy gap. These pieces of evidence need to be reconciled so as to shed light on the physics of $MgB_2$.

We have performed a systematic study of point-contact tunnelling spectra obtained on high-quality $MgB_2$ samples. We confirm that the two-gap state is intrinsic to $MgB_2$, although the high-energy gap is observable only under certain experimental conditions.

## Technique

The metal-superconductor tunneling spectra were measured using a point contact technique. The contacts were established by driving a sharpened gold tip into the flat sample surface. The tip was mounted on a pin attached to a differential screw. The screw was operated by means of a long shaft from the top of the cryogenic insert. The vertical movement of the tip could be controlled with a nominal precision of about 0.5 micron, and so that forces in the gram-range could be exerted on the contact. The sample was mounted on a sliding stage, thus providing *in situ* horizontal sample movement. The advantage of this design is that it essentially eliminates the effects of thermal contraction, so that in many cases it was possible to vary the junction temperature without affecting the contact.

The tips were normally made of 0.5-mm gold wire sharpened so that the apex size was originally of order of 10 - 20 microns. To establish a contact, the tip was pushed into the sample surface and squashed against it, so that the actual contacts were some 50 microns in diameter. Other tip materials were tried as well. In terms of the principal tunneling features, the results obtained with different tip metals were very similar, but as gold is the softest, it avoids damage to the sample surface.

The measurement protocol was as follows: The insert was placed into a liquid-helium dewar with the tip separated from the sample, and the contact established with the sample at the helium temperatures. We have found that junctions have good temporal and mechanical stability once the resistance is brought to around 100 Ohms. Starting from that point, the junction differential conductance was measured as a function of applied voltage, using a standard lock-in technique. At a given location on the sample surface, the datasets were taken as the background junction conductance $\sigma_0$ was varied (by driving the tip harder into the surface), or the temperature evolution of spectra at a particular $\sigma_0$ could be followed. The typical range of resistances was between few Ohms and few hundreds Ohms; the temperature was stabilised to within 0.1 K. After the data had been taken at a particular point on the sample, the tip was lifted so as to break the contact, and the sample moved; the measurements were then repeated at a new location. If the insert had to be warmed up in order to examine or replace the sample, the tip was replaced as a matter of course.

## Samples

Most of our measurements were made on a 39-K $MgB_2$ thin film grown by electron-beam evaporation, followed by reaction with magnesium [17]. Scanning-electron microscope images (Fig. 1) show that the film is dense, although granular; the grains are approximately 100-500 nm in diameter. The film, as opposed to pulsed-laser-deposited (PLD) $MgB_2$ films [18], was fabricated without the presence of substantial MgO in the target. Over most of the surface the grains are barely visible, so that it is the surface roughness rather than intergranular cracks and voids that give an idea of the grain dimensions. The good grain connectivity accounts for the low resistivity of this film, 18 μOhm·cm at 40 K. The superconducting transition is sharp, the midpoint is at 38.6 K, and the width (determined as the interval between 10% and 90% of the normal-state resistance), is 0.6 K.

Some spectra were obtained on the less-homogeneous PLD films [18], and on polycrystalline fragments extracted from the commercial $MgB_2$ powder.

No special measures were taken to prepare the film surfaces before the experiments, apart for cleaning them in ethyl alcohol. The polycrystalline fragments were polished on fine diamond paste to produce a flat surface. We assume that low tunneling barriers in the point contacts stemmed from the naturally slightly degraded surface layer.

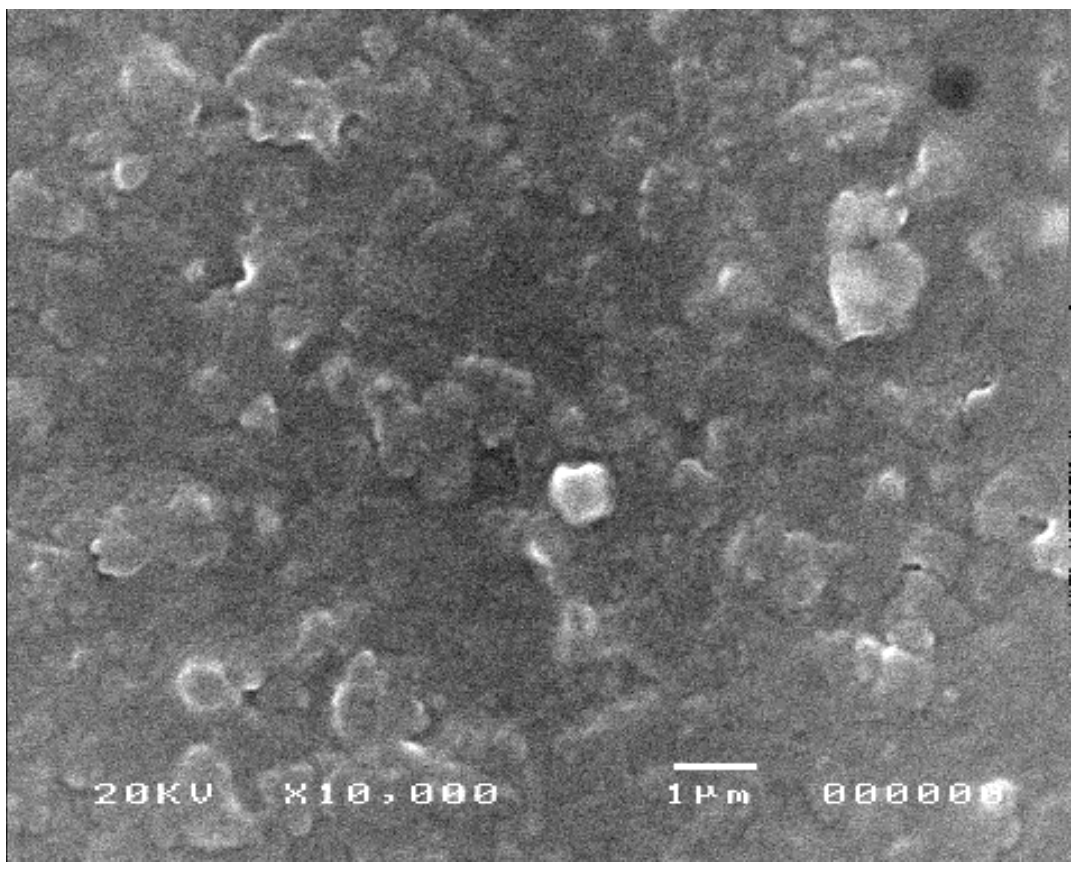

Fig.1. SEM image of the $MgB_2$ thin film.

As we can judge from the relative sizes of the grains and the tip, there was a large number of grains sampled simultaneously in the contact area. The clearness and reproducibility of spectra indicate a good degree of local (on the scale of 50 μm) homogeneity of the samples. From the analysis of the results, we have got an indication that the individual metal-superconductor junctions within the contact were small compared to the grain size; we therefore assume that a net measured spectrum represents a sum of spectra stemming from tunneling into individual grains.

## *Theoretical background and the fitting procedure.*

Tunneling spectroscopy is a direct method of measuring the single-particle excitation spectra in superconductors [19]. It comprises measurement of the differential conductivity of a junction between a superconductor and a normal metal. If there is a dielectric layer between the two electrodes (N-I-S contact), the conductance as a function of applied dc bias across the junction yields the quasiparticle density of states in the superconductor. The conductance at energies within the gap is zero due to absence of the quasiparticle states; at voltages corresponding to $\pm\Delta/e$ sharp peaks appear ("coherence peaks"), which are a readily detectable experimental indication of the gap. When the insulating barrier is removed and instead a direct N-S junction is formed, there appears a non-trivial conduction channel at small energies, as a result of the Andreev reflection process . In this regime the conductance within the gap is twice as high as the background conductance. The intermediate situation of a partially-transparent barrier was solved analytically by Blonder, Tinkham and Klapwijk (BTK) [20].

For a single-gap spectrum there are three parameters in the model that are varied to obtain a fitting curve: the gap $\Delta$, the barrier strength Z and the effective smearing parameter $\Gamma$. Pure Andreev reflection corresponds to Z=0, whereas in the N-I-S tunneling regime $Z \to \infty$. There are several different physical mechanisms that cause smearing: a finite quasi-particle lifetime, finite temperature, and inhomogeneity of the material properties in the junction area. In fitting, we made no *a priori* assumptions as to the nature of the smearing, but rather determined a single effective parameter $\Gamma_{eff}$, which incorporates the overall effect of all the mechanisms. For simplicity, the effect of smearing was taken into account by convolution of the zero-temperature function with a gaussian distribution. To give an example of the effect of the smearing parameter on a tunneling spectrum, in Fig. 2 we show a smeared function ($\Gamma$=0.9 meV or approximately 10 K) as compared to the zero-temperature curve. The value of Z is the same for the two curves, Z=0.5.

Depending on the ratio between the junction size *d* and the mean-free path *l* in the superconductor, the junction can be in either the ballistic ($d < l$) or the diffusive ($d > l$) regime. The two cases were considered theoretically by Mazin *et al.* [21]. Our fitting results suggest that the data are consistent with the ballistic regime of Andreev reflection. The mean-free path in $MgB_2$ is of the order of 10 nm [22], therefore it is likely that there is a large number of junctions in the 50-μm contact area. Since they are all connected in parallel, the total conductance is the sum of the conductances of the individual junctions.

When spectra with a double gap structure were fitted, it was assumed that they were composed of two independent current channels, taken with appropriate weights, $\sigma(V) = (1-f)\sigma_S(V) + f\sigma_L(V)$, where $\sigma_S$ and $\sigma_L$ are the contributions corresponding to the small and large gaps, respectively. When performing such fits, we assumed that the barrier strength and the smearing parameter were the same for both contributions, in order to minimise the number of free parameters.

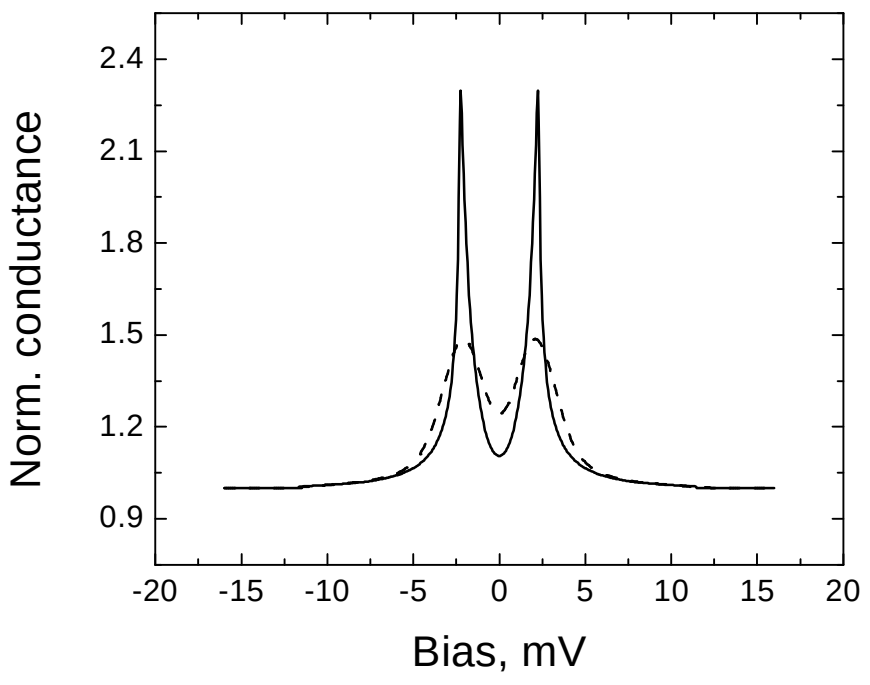

Fig. 2 The effect of smearing on simulated spectra for a gap of 2.3 meV: Γ=0 (solid line) and Γ=0.9 meV (dashed line). Z=0.5 for both curves.

## *Experimental results*

### a) Varying the background conductance using contact pressure.

Examples of the tunneling spectra of the $MgB_2$ thin film at 4.2 K are shown in Fig. 3. The bottom curves of these graphs were the first to be taken; the successive data sets correspond to increasing contact pressure. The structure becomes more pronounced as the background conductance increases, and the presence of two sets of peaks, at about ±2.3 meV and about ±6.5 meV is apparent in Fig. 3a. On the other hand, a number of measurements taken on the same film, but at different locations, revealed data of similar quality, but showing only the low-voltage feature (Fig 3b). Analogous measurements performed on a polycrystalline fragment extracted from commercial $MgB_2$ powder yielded single-gap structures as well (Fig. 3c).

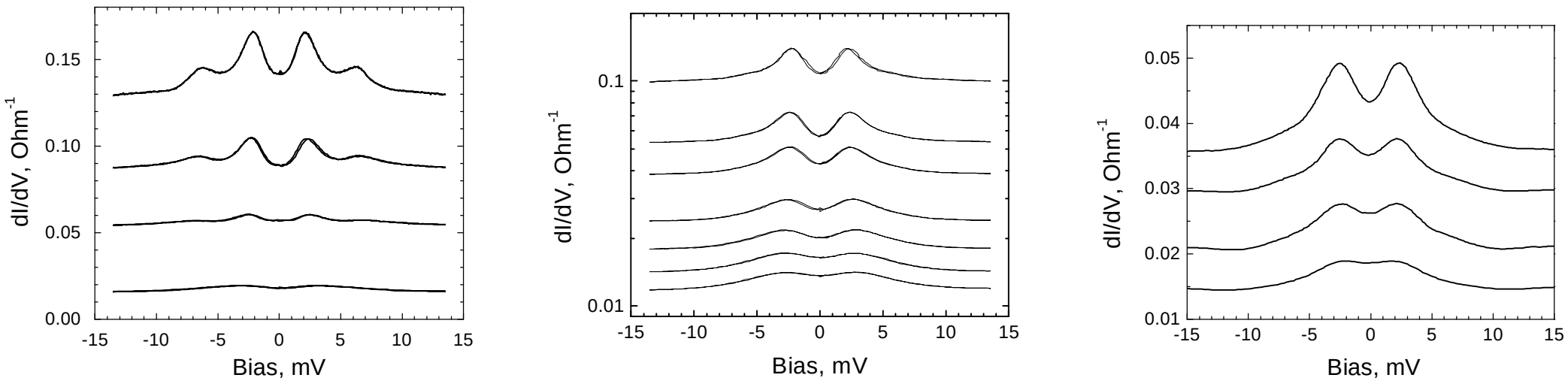

(a) (b) (c)

Fig. 3. Variation of the spectra with changing the contact pressure at 4.2 K. a) Thin film, double feature evident. b) Spectra at another location on the same film; no visible double structure. c) Polycrystalline fragment, low-energy feature dominant In each experiment, the low-conductance (low contact pressure) data were taken first, then the pressure was progressively increased.

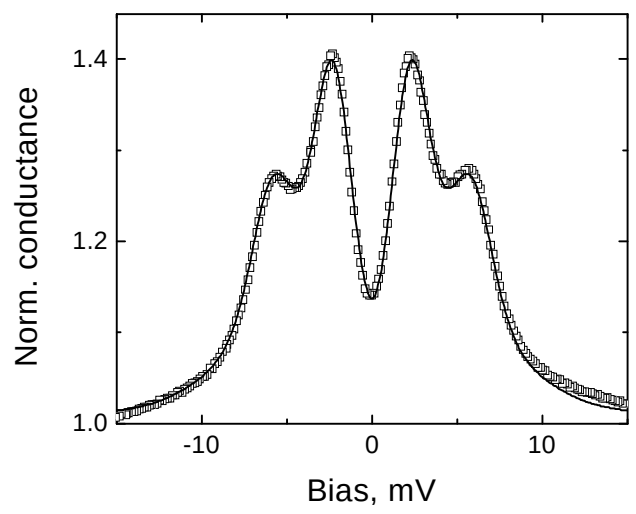

Fig. 4. An example of fitting to the BTK dependence for a $MgB_2$ thin film, T=4.2 K. Points: experimental data; line: the fit taking the ballistic regime, $\Delta_S$ = 2.4 meV, $\Delta_L$ =6.2 meV, Z = 0.53, weight of the high-energy contribution f = 0.28.

The majority of the curves were successfully fitted to the smeared BTK function, i.e. the junctions were in the ballistic limit. The agreement between the data and the fitting function was excellent, as illustrated in Fig. 4. Analogous measurements and fitting were carried out on a number of PLD films. Thus, good statistics have been collected.

The low-energy gap was apparent in all the spectra; its values group around $\Delta_S$ = 2.3 meV with a statistical spread of 0.3 meV. The position of the larger gap feature, at $\Delta_L$ = 6.2±0.7 meV, was determined with a slightly poorer accuracy, as its contribution was often obscured by the dominant low-energy structure. The barrier height, Z, decreased systematically with increasing junction background conductance $\sigma_0$, as it should (Fig. 5a). A less trivial observation is that the effective smearing parameter, Γ, shows a strong dependence on $\sigma_0$ (Fig. 5b). This may indicate that the disorder or carrier scattering in the barrier contribute to Γ. Had Γ been determined by temperature smearing only, its value should be close to 0.36 meV, which is equivalent to 4.2 K. As shown in Fig. 5b, it is only for the highest values of $\sigma_0$ that Γ approaches this value. The decreasing Γ for more transparent barriers implies that the disorder becomes less important as the barrier is thinner, which seems plausible.

The key observation concerns the evolution of the high-energy feature. The relative weight of its contribution systematically increases with increasing $\sigma_0$ (Fig. 5c). The data scatter in this graph is substantial, as we have found that even on the same film the details of spectra varied at different locations. Therefore, while it would be inappropriate to extract definitive functional dependences from the data in Figs 5a-5c, the general trends are apparent. In what follows we incorporate these trends into the interpretation of the results.

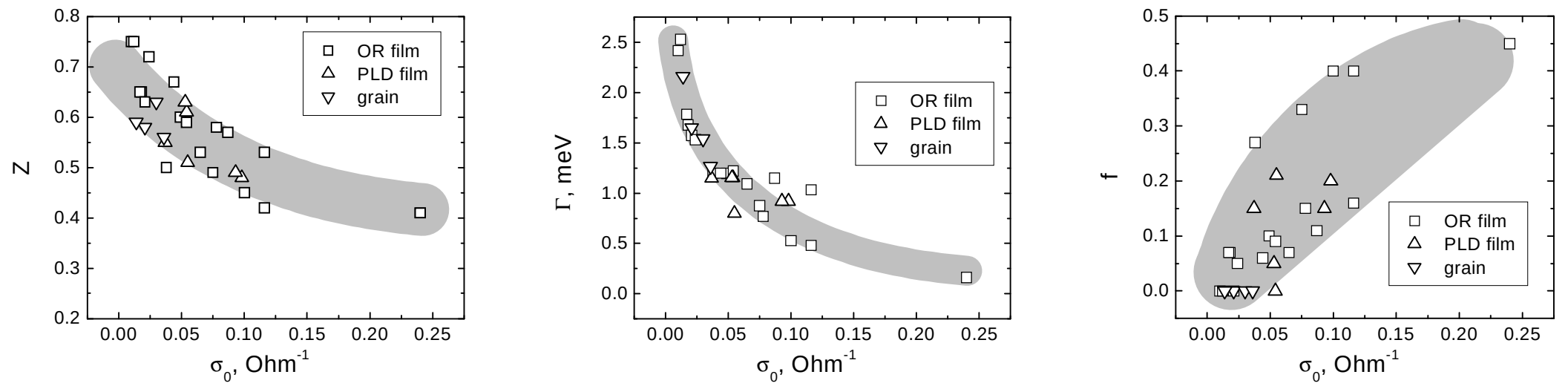

(a) (b) (c)

Fig. 5. Variation of the barrier height, Z, the smearing parameter, Γ, and the weight of the higher-energy feature, f, as a function of the background conductance of the junction. All the data are at 4.2 K.

## b) Temperature evolution of spectra

The low-resistance junctions were stable enough that it was possible to take temperature dependences of the spectra while the junction was not altered mechanically. This junction stability was manifested by the constancy of the background conductance as the temperature was varied. Fig.6 presents the temperature evolution of a contact with a simpler case of a single-gap spectrum. It is noteworthy that, although even for this simple case there are three fitting parameters, they appear to be 'orthogonal' in the sense that each of them is sensitive to a particular aspect of the spectrum: At low temperatures, the position of the peaks defines Δ to within 0.2 meV, Γ is related to the peak width, and Z determines the amplitude of the peaks and the depth of the central dip. The fit remains unique and well-defined at higher temperatures as well. The dip completely smears out at around 17 K (equivalent to 1.5 meV on the energy scale), so that the individual peaks are not resolved at higher temperatures. However, a plateau is retained at the spectrum centre up to 25 K, as a direct indication of the gap value. At higher temperatures, the fit becomes less unambiguous, so we had to fix the value of Z, allowing the other two parameters to vary. Above 32 K the spectra became too smeared, and, although the peak survived until the temperature exactly reached $T_c$, no attempt was made to fit the data in this range.

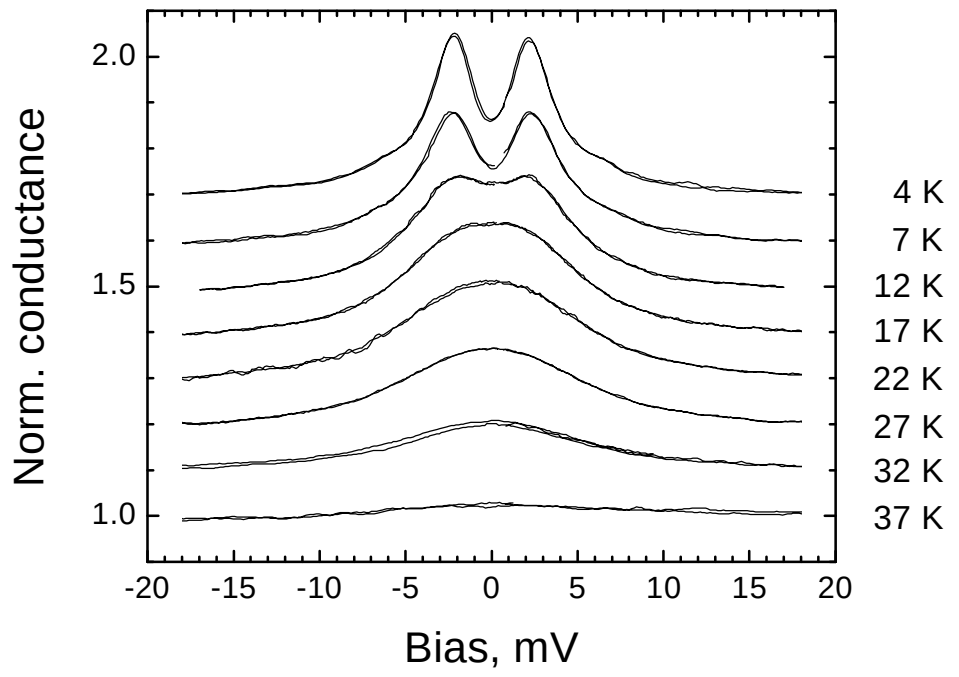

Fig. 6. Temperature evolution of a single-gap spectrum, starting from 4.2 K and increasing T subsequently. Curves are shifted for clarity.

The first proof of the validity of the fitting procedure is the behaviour of the barrier height: as the temperature varied, Z stayed constant unless an obvious jump in the background conductance occurred. The smearing parameter increased in one-to-one relation with temperature (Fig. 7), although there was a positive offset. The fact that, apart from this offset, Γ is equal to the measurement temperature, is further strong evidence that the fitting was valid. The temperature dependence of the gap inferred from this analysis is shown in Fig. 8. It clearly shows that the small gap, when observed alone, closes at the bulk $T_c$ of $MgB_2$, and therefore is intrinsic rather than related to a spurious low-$T_c$ phase.

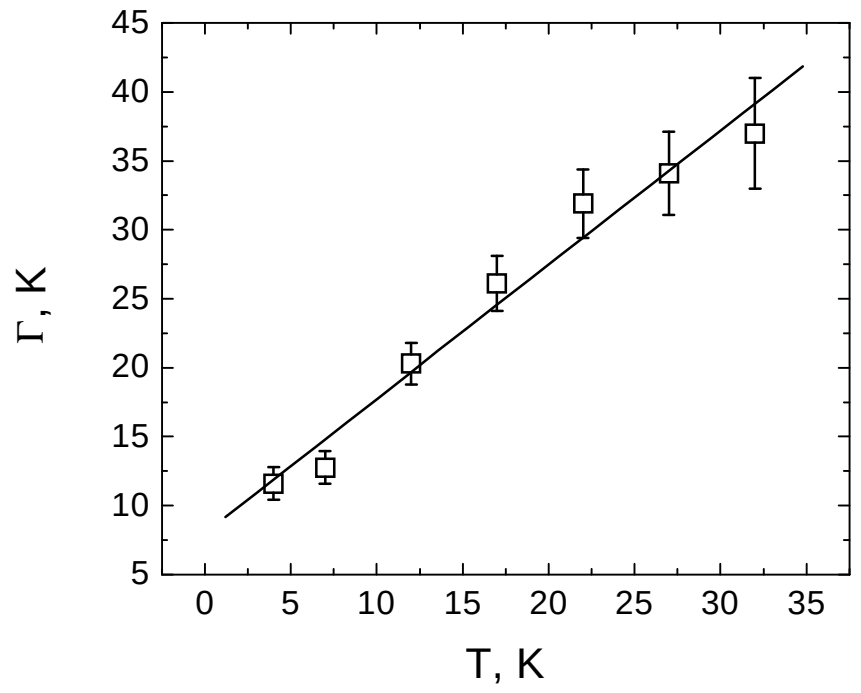

Fig.7. The smearing parameter Γ, expressed in Kelvin, as a function of temperature. The line is the best linear fit; its slope is 0.95.

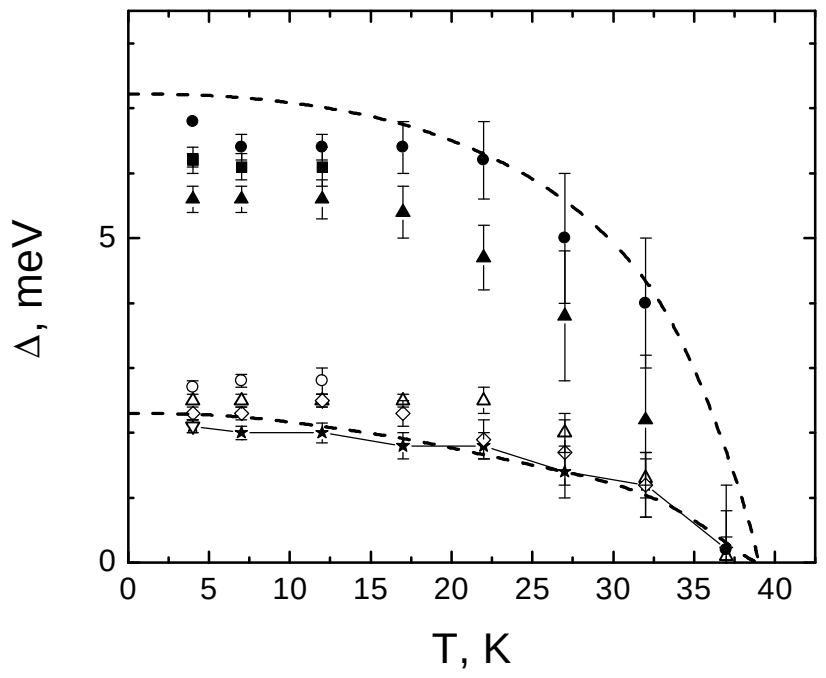

Fig.8. The temperature dependences of the gaps, as inferred from the fitting procedure. The data points connected by line segments correspond to single-gap spectra. Other points are inferred from the spectra in which both gaps were present simultaneously. Dashed lines are the results of calculations in Ref [25].

The rest of the points in Fig. 8 were obtained from analysis of spectra where the double structure was evident at low temperatures. The relevance of the two-gap fitting was beyond question below 15 K, as two pairs of peaks were then visible. At higher temperatures the structure smears into a single broad peak, so that the extraction of two distinct values of the gap might be more doubtful. The convincing evidence is shown in Fig.9: single-gap fitting fails with either of the small or large gaps alone. In contrast, the combined fit agrees with the data very well. The parameters Z and f are the same as at other temperatures, and $\Gamma$ agrees with the temperature of measurement. The single-gap function can be forced to fit these data, but only at the expense of varying Z and setting $\Delta$ at 2.5 meV, higher than the small gap at low temperature. Therefore the two-gap fitting is reliable, and thus confirms the presence of both features at high temperatures.

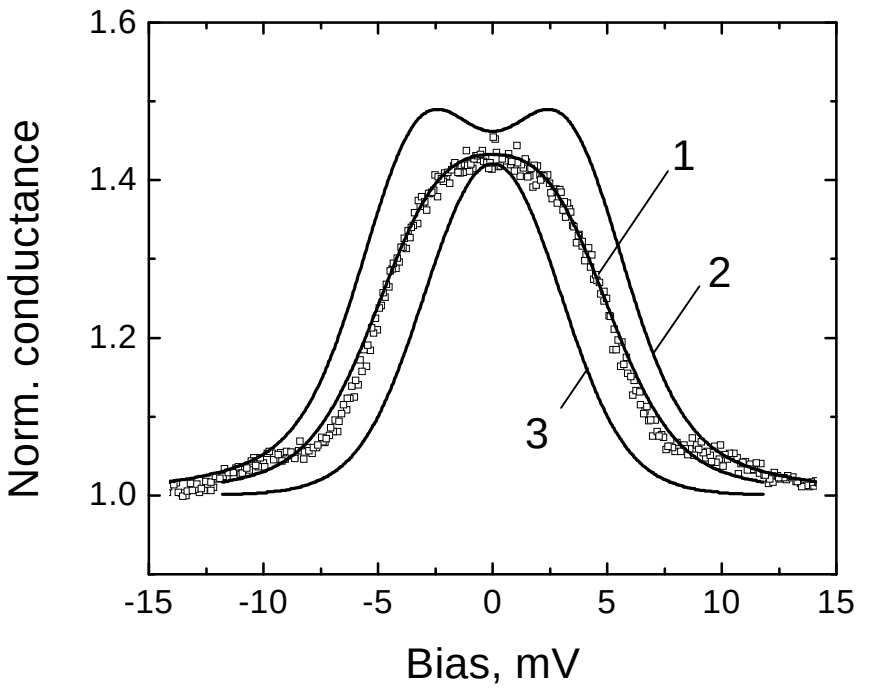

Fig. 9. Data (points) and fitting functions (curves) at 27 K. A double-gap fit (curve 1, $\Delta_1$=1.7, $\Delta_2$=3.8, f=0.55) is in good agreement with the data; single-gap fits with the previously-determined gaps (curve 2 large gap only; curve 3 small gap only), are clearly incompatible with the experiment. The best fit to a single-gap function (not shown) works worse than the double-gap, and yields $\Delta$=2.5 meV, too large for the small gap at this temperature, and too small for the large gap.

## *Discussion*

Our results are in line with the results of Szabo et al. [8], and make the observations of single anomalously small gaps [3, 11] comprehensible. Although the two gaps are intrinsic, whether they are sufficiently resolved to be visible depends on the barrier transparency, and so on the contact pressure. As we have observed, the weight of the large gap contribution increases as the barrier transparency is increased; in other words, a redistribution of current between the two channels occurs. This fact suggests that the dependence of the tunneling probability on contact pressure is different for the two channels. At low pressures (high barriers) the effective conductance of the large-gap channel is much less than that of the small-gap channel. At low barriers, the conductances become nearly equal. In very general terms, the effective barrier transparency depends on, among other parameters, the ratio between the geometric thickness of the insulating (oxide) layer to the superconducting coherence length $\xi$: the smaller is $\xi$, the lower the transparency. In a clean superconductor, $\xi$ is inversely proportional to $\Delta$ [23]: $\xi = 2\hbar v_F / \pi\Delta$ (where $v_F$ is the Fermi velocity).

Let us assume for simplicity (and supported by our experimental observations) that there are only two different wavefunctions, with distinctly different gap values, and also let us assume that the material is in the clean limit, (as many results suggest, e.g. [22]), then it is valid to conclude that there will be two correspondingly different coherence lengths also. In this case the barrier of a certain geometrical thickness *t* has effectively high transparency for the small-gap, long $\xi$ channel, and low transparency for the other channel. As *t* is decreased (by increased contact pressure ) to the point when it is comparable with the shorter $\xi$, the conductances of the two channels will tend to equalise.

This argument explains the main trend in the data of Fig.5c. However, on top of this trend, there is a significant scatter of data points from measurements taken in different locations on the same film. This suggests that contact pressure is not the only variable that determines the relative amplitude of the large-gap feature. Indeed, given that the material is also anisotropic, the local degree of grain alignment will also contribute to the transmission probability of each channel 24. In this way granular randomly-oriented or completely grain-aligned samples will be expected to have different contact-pressure dependent weighting factors for the two gap features.

The sizes of the gaps and their temperature dependences, are in good agreement with the calculations by Liu *et al.* [25]. According to ref [25], the Fermi surface of $MgB_2$ comprises multiple sheets: two nearly cylindrical hole sheets arising from quasi-2D $p_{x,y}$ boron bands, and two tubular networks arising from 3D $p_z$ bonding and anti-bonding bands. Using these calculations the superconducting state can then be described by two distinct gaps opening at the 2D and 3D Fermi surfaces. The calculations of phonon properties within this scenario make it possible to account for a transition temperature of about 40 K. The symmetry of two gaps follows the symmetry of the Fermi surface. The value of the 2D gap is about 7.2 meV, which is higher than isotropic BCS theory would yield, and the 3D gap is about 2.3 meV. These numbers agree with the experiment. Moreover, as the larger gap is 2-dimensional, it is only observable when the tunneling direction is parallel to the *ab* plane of crystallites. The smaller 3-dimensional gap is not sensitive to the tunneling direction. We see from this theory that when the $MgB_2$ sample is granular and randomly-oriented, the probability of observing the large gap structure is low in proportion to the fraction of crystallites with *ab* planes close to the tunneling direction. If there is a variation of texture across the sample surface, the probability of observing the large gap would also vary depending on the location. Our films were partially c-axis aligned [17], hence within this scenario it is likely that the degree of local texture contributes to the scatter in the large gap amplitude (at fixed background conductance).

The non-thermal smearing increases with increasing the effective barrier height in a consistent manner. There are no indications that the associated gap values decrease, as one would expect if tunneling current were flowing into regions with suppressed superconductivity. It is plausible that it is the scattering in the less-transparent (or thicker) barrier that causes the spectra to smear. Thus our observations are compatible with some of the other experiments, and with the theoretical predictions by Liu et al. [25]. Nevertheless, it has to be stressed that these results in themselves do not contain any unambiguous evidence as to the symmetry or dimensionality of the order parameter. Therefore it would be wrong at this juncture to rule out other possible scenarios. The strongest alternative candidates for an explanation of two distinct gaps in $MgB_2$ are: 1) anisotropic s-wave order parameter yielding different experimental results according the direction of the current [26]; and 2) non-trivial surface superconducting states showing effectively depressed gap [27]. It seems unlikely that the issue can be fully resolved until experiments have been done on very well aligned (ideally, single-crystalline) samples of $MgB_2$.

## Conclusions

The main conclusions that we draw are as follows:

1. There are two distinct contributions to the tunneling spectra of $MgB_2$/gold point contacts, corresponding to gaps (as T tends to 0) of 2.3 meV and 6.5 meV. They must be intrinsic to $MgB_2$, because both gaps close at the bulk $T_c$ of the material.
2. The relative contributions of the two channels vary for different barrier transparencies. When the overall junction resistance is high, the low-gap channel is dominant. As the junction resistance is decreased, the contribution of the larger gap increases. The low gap remains in place throughout the whole range of resistances. This fact can be explained by different gap states having different coherence lengths . In the clean limit, the lower-gap wave function has a larger $\xi$, and therefore an oxide layer of a certain thickness is more transparent to this function rather than to the large-gap one.
3. There is significant temperature-independent smearing of the spectra. The smearing increases as the barrier becomes thicker, and is therefore likely to be caused by the disorder or quasiparticle scattering inside the barrier.

## Acknowledgements

We are grateful to Dr Mike Osofsky (NRL, Washington) for useful discussions. This work was supported by the UK Engineering and Physical Sciences Research Council.